\documentclass[aps, pra, twocolumn,showpacs,preprintnumbers,amsmath,amssymb]{revtex4}
\newcommand{\ket}[1]{\mbox{$ | #1 \rangle $}}

\usepackage[dvips]{graphicx}% Include figure files

\begin{document}

\preprint{}

\title{Quantum secure direct communication with pure entangled states}
% Force line breaks with \\

\author{Jian Wang}

 \email{jwang@nudt.edu.cn}

\affiliation{School of Electronic Science and Engineering,
\\National University of Defense Technology, Changsha, 410073, China }
\author{Quan Zhang}
\affiliation{School of Electronic Science and Engineering,
\\National University of Defense Technology, Changsha, 410073, China }
\author{Chao-jing Tang}
\affiliation{School of Electronic Science and Engineering,
\\National University of Defense Technology, Changsha, 410073, China }

 %\email{ishizuka@isl.melco.co.jp}
% \affiliation{School of Electronic Science and Engineering, \\National Univ of
%Defense Technology, Changsha , 410073,China}
%{Information Technology R \& D Center, Mitsubishi Electric
%Corporation\\
%5-1-1 Ofuna, Kamakura, Kanagawa 247-8501, JAPAN\\
%TEL: +81-467-41-2190 \quad FAX: +81-467-41-2185}


\begin{abstract}
We present a quantum secure direct communication protocol where the
channels are not maximally entangled states. The communication
parties utilize decoy photons to check eavesdropping. After ensuring
the security of the quantum channel, the sender encodes the secret
message and transmits it to the receiver by using Controlled-NOT
operation and von Neumann measurement. The protocol is simple and
realizable with present technology. We also show the protocol is
secure for noisy quantum channel.
\end{abstract}

\pacs{03.67.Dd, 03.65.Ud, 42.79.Sz}% PACS, the Physics and Astronomy
\keywords{Quantum secure direct communication; Pure entangled
states}
%Use showkeys class option if keyword
\maketitle

%%%%%%%%%%%%%%%%%%%%%%%%%%%%%%%%%%%%%%%%%%%%%%%%%%%%%%%%%%%%%%%%%%%%%
%=====================================================================
In recent years, a novel concept, quantum secure direct
communication (QSDC) has been proposed \cite{beige}. Different from
quantum key distribution whose object is to establish a common key
between the communication parties, QSDC's object is to transmit the
secret message directly without first establishing a key to encrypt
it. QSDC can be used in some special environments, which has been
shown in Ref. \cite{Bostrom,Deng}. The works on QSDC attracted a
great deal of attention. We can divide these works into two kinds,
one utilizes single photon \cite{denglong,cai1,jwang1}, the other is
based on entangled state
\cite{Bostrom,Deng,cai4,jwang2,jwang3,cw1,cw2,yan,man,gao1,gao2}.
Deng et al. \cite{denglong} proposed a QSDC protocol by using
batches of single photons which serves as quantum one-time pad
cryptosystem. Cai et al. \cite{cai1} presented a deterministic
secure direct communication protocol using single qubit in a mixed
state. We proposed a QSDC protocol based on the order rearrangement
of single photons \cite{jwang1}. The QSDC protocol using
entanglement state is certainly the mainstream. Bostr\"{o}m and
Felbinger \cite{Bostrom} proposed a Ping-Pong protocol which is
quasi-secure for secure direct communication if perfect quantum
channel is used. Cai et al. \cite{cai2,cai3} pointed out that the
Ping-Pong protocol is vulnerable to denial of service attack or
joint horse attack with invisible photon. They also presented an
improved protocol which doubled the capacity of the Ping-Pong
protocol \cite{cai4}. Zhang et al. \cite{zhang05} indicated that the
ping-pong protocol can be eavesdropped even in an ideal quantum
channel. Deng et al. \cite{Deng} put forward a two-step QSDC
protocol using Einstein-Podolsky-Rosen (EPR) pairs. We presented a
multiparty controlled QSDC protocol by using
Greenberger-Horne-Zeilinger (GHZ) state and a QSDC protocol without
using perfect quantum channel \cite{jwang2,jwang3}. Wang et al.
\cite{cw1,cw2} proposed a QSDC protocol with quantum superdense
coding and a multi-step QSDC protocol with a sequence of
Greenberger-Horne-Zeilinger states. Yan and Zhang \cite{yan}
presented a QSDC protocol using EPR pairs and teleportation. Man et
al. \cite{man} proposed a QSDC protocol by using swapping quantum
entanglement and local unitary operations. Gao et al.
\cite{gao1,gao2} presented a QSDC protocol by using GHZ states and
entanglement swapping and a controlled QSDC protocol using
teleportation.

Many QSDC protocols require maximally entangled states. However, we
do not have maximally entangled states because of decoherence and
noise. Moreover, it can only obtain almost maximally entangled
states from partially entangled states by using quantum
distillation. In this paper, we present a QSDC protocol using pure
entangled states. In the protocol, the communication parties utilize
decoy photons to ensure the security of the quantum channel, which
is similar to the method used in Ref. \cite{li}. If there is no
eavesdropping in the transmission line, the sender encodes the
secret message and transmits it to the receiver by using
Controlled-NOT (CNOT) operation and von Neumann measurement.
According to the sender's measurement result, the receiver can
obtain the secret message. In the protocol, the transmitting photon
sequence does not carry the secret message. We also show the present
protocol is secure even with a noisy channel.

Suppose the sender, Alice, wants to transmit her secret message to
the receiver, Bob directly. The details of our QSDC protocol is as
follows:

(S1) Alice prepares an ordered $N$ two-photon states, each of
which is in the state
\begin{eqnarray}
\ket{\phi}_{AB}=a\ket{00}_{AB}+b\ket{11}_{AB},
\end{eqnarray}
where $|a|^2+|b|^2=1$. We denotes the ordered $N$ states with
\{[P$_1(A)$,P$_1(B)$], [P$_2(A)$,P$_2(B)$], $\cdots$,
[P$_N(A)$,P$_N(B)$]\}, where the subscript indicates the pair
order in the sequence, and $A$ and $B$ represent the two photons
of each state. Alice takes one photon from each state to form an
ordered partner photon sequence [P$_1(A)$, P$_2(A)$,$\cdots$,
P$_N(A)$], called $A$ sequence. The remaining partner photons
compose $B$ sequence, [P$_1(B)$, P$_2(B)$,$\cdots$, P$_N(B)$].

(S2) Alice prepares some decoy photons for eavesdropping check.
Each of the decoy photons is randomly in one of the four states
\{\ket{0}, \ket{1}, \ket{+}=$\frac{1}{\sqrt{2}}(\ket{0}+\ket{1})$,
\ket{-}=$\frac{1}{\sqrt{2}}(\ket{0}-\ket{1})$\}. Alice inserts the
prepared decoy photons in $B$ sequence randomly and sends the new
$B$ sequence to Bob.

(S3) After confirming that Bob has received $B$ sequence, Alice
publishes the position and basis of the decoy photons in $B$
sequence. Bob performs von Neumann measurement on the decoy photons
according the corresponding measuring basis and announces publicly
his measurement result. According to Bob's result, Alice can then
analyze the error rate during the transmission of $B$ sequence. If
the error rate is below the threshold they preset, Alice can
conclude that there is no eavesdropper in the line. Alice and Bob
continue to the next step. Otherwise, they abort the communication.

(S4) Alice prepares a photon $a$ according to the bit value of her
secret message. If Alice's secret message bit is ``0''(``1''), she
prepares a photon $a$ in the state $\ket{0}$ ($\ket{1}$). Thus Alice
prepares $N$ photons for the ordered $N$ states, which we call $a$
sequence, [P$_1(a)$, P$_2(a)$,$\cdots$, P$_N(a)$]. If the state of
photon P$_i(a)$ ($i=1,2,\cdots,N$) is $\ket{0}$, then the state of
the system composed by photons P$_i(a)$, P$_i(A)$ and P$_i(B)$ is
\begin{eqnarray}
\ket{\Phi_0}_{aAB}=\ket{0}_a\otimes(a\ket{00}+b\ket{11})_{AB},
\end{eqnarray}
where the subscript $a$ denotes P$_i(a)$. If the state of P$_i(a)$
is $\ket{1}$, then the state of [P$_i(a)$, P$_i(A)$, P$_i(B)$] is
\begin{eqnarray}
\ket{\Phi_1}_{aAB}=\ket{1}_a\otimes(a\ket{00}+b\ket{11})_{AB}.
\end{eqnarray}

(S5) Alice sends photons P$_i(A)$ and P$_i(a)$ ($i=1,2,\cdots,N$)
through a CNOT gate (photon P$_i(A)$ is the controller and photon
P$_i(a)$ is the target). Thus $\ket{\Phi_0}_{aAB}$ is changed to
\begin{eqnarray}
\label{7}
\ket{\Phi_0'}_{aAB}=(a\ket{000}+b\ket{111})_{aAB}
\end{eqnarray}
and $\ket{\Phi_1}_{aAB}$ becomes
\begin{eqnarray}
\label{8}
\ket{\Phi_1'}_{aAB}=(a\ket{100}+b\ket{011})_{aAB}.
\end{eqnarray}

(S6) After having done CNOT operation, Alice and Bob measures photon
P$_i(a)$ and P$_i(B)$ in $Z$-basis, $\{\ket{0}, \ket{1}\}$,
respectively. According to equations (\ref{7}) and (\ref{8}),
although Bob obtains his measurement result, he cannot recover the
secret message without Alice's result.

(S7) Alice publishes her measurement results of the photons in $a$
sequence. Referring to Alice's result, Bob can recover Alice's
secret message, as illustrated in Table 1.
\begin{table}[h]
\caption{The recovery of Alice's secret message }\label{Tab:one}
  \centering
    \begin{tabular}[b]{|c|c|c| c|} \hline
      Alice's result & Bob's result & secret message\\ \hline
      \ \ket{0} & \ket{0} & 0\\ \hline
       \ \ket{0} & \ket{1} & 1\\ \hline
        \ \ket{1} & \ket{0} & 1\\ \hline
         \ \ket{1} & \ket{1} & 0\\ \hline
        \end{tabular}
\end{table}
For example, when Bob's result is $\ket{0}$, he can conclude that
the Alice's secret message is ``0'' (``1''), if Alice's result is
\ket{0} (\ket{1}).

So far we have proposed the QSDC protocol with pure entangled
states. Now, let us discuss the security for the present protocol.
The crucial point is that the inserted decoy photons does not allow
an eavesdropper, Eve, to have a successful attack and Eve's attack
will be detected by the communication parties during the
eavesdropping check. Because each of the decoy photons is randomly
in one of the four states \{\ket{0},\ket{1},\ket{+},\ket{-}\}, the
security for the protocol is the same as that for BB84 protocol
\cite{bb84}. If the security of the quantum channel is ensured, the
protocol is completely secure.

According to Stinespring dilation theorem \cite{Bostrom}, Eve's
attack can be realized by a unitary operation $\hat{E}$ on a large
Hilbert space, $H_{AB}\otimes H_{E}$. The state of decoy photon and
Eve's probe state is
\begin{eqnarray}
\hat{E}\ket{0,\varepsilon}=\alpha\ket{0,\varepsilon_{00}}+\beta\ket{1,\varepsilon_{01}},\\
\hat{E}\ket{1,\varepsilon}=\beta'\ket{0,\varepsilon_{10}}+\alpha'\ket{1,\varepsilon_{11}},
\end{eqnarray}
\begin{eqnarray}
\hat{E}\ket{+,\varepsilon}&=&\frac{1}{\sqrt{2}}[\alpha\ket{0,\varepsilon_{00}}+\beta\ket{1,\varepsilon_{01}}\nonumber\\
&+&\beta'\ket{0,\varepsilon_{10}}+\alpha'\ket{1,\varepsilon_{11}}],\nonumber\\
&=&\frac{1}{2}[\ket{+}(\alpha\ket{\varepsilon_{00}}+\beta\ket{\varepsilon_{01}}+\beta'\ket{\varepsilon_{10}}+\alpha'\ket{\varepsilon_{11}})\nonumber\\
&+&\ket{-}(\alpha\ket{\varepsilon_{00}}-\beta\ket{\varepsilon_{01}}+\beta'\ket{\varepsilon_{10}}-\alpha'\ket{\varepsilon_{11}})],\\
\hat{E}\ket{-,\varepsilon}&=&\frac{1}{\sqrt{2}}[\alpha\ket{0,\varepsilon_{00}}+\beta\ket{1,\varepsilon_{01}}\nonumber\\
&-&\beta'\ket{0,\varepsilon_{10}}-\alpha'\ket{1,\varepsilon_{11}}],\nonumber\\
&=&\frac{1}{2}[\ket{+}(\alpha\ket{\varepsilon_{00}}+\beta\ket{\varepsilon_{01}}-\beta'\ket{\varepsilon_{10}}-\alpha'\ket{\varepsilon_{11}})\nonumber\\
&+&\ket{-}(\alpha\ket{\varepsilon_{00}}-\beta\ket{\varepsilon_{01}}-\beta'\ket{\varepsilon_{10}}+\alpha'\ket{\varepsilon_{11}})],
\end{eqnarray}
where \ket{\varepsilon} denotes Eve's probe state. The probe
operator can be written as
\begin{eqnarray}
\hat{E}=\left( \begin{array}{c c} \alpha &\;\;  \beta' \\
\beta &\;\;  \alpha'
\end{array} \right).
\end{eqnarray}
As $\hat{E}$ is an unitary operation, the complex numbers $\alpha$,
$\beta$, $\alpha'$ and $\beta'$ must satisfy $\hat{E}\hat{E}^\dag=I$
and we can obtain the relations
\begin{eqnarray}
|\alpha'|^2=|\alpha|^2, |\beta'|^2=|\beta|^2.
\end{eqnarray}
The error rate introduced by Eve is $e=|\beta|^2=1-|\alpha|^2$.

The above analysis is based on ideal circumstances and does not take
into account noise in the transmission line. In noisy quantum
channel, Eve intercepts some transmitting photons in $B$ sequence at
step (S2) and sends the others to the receiver using a better
quantum channel in which the photon loss will not increase. During
the eavesdropping check, Eve's attack will not be detected in this
situation. However, according to our protocol, Eve cannot obtain
Alice's secret message without Alice's measurement result even if
she captures some photons in $B$ sequence. If the eavesdropping
check is passed, Bob tells Alice which photon he has received and
which photon is lost in the transmitting line at step (S6) of the
protocol. Alice then only publishes her measurement results of the
corresponding photons which Bob has received. For example, if Bob
has received photons $P_3(B)$, $P_6(B)$, $\cdots$ in $B$ sequence,
then Alice publishes her measurement results of photons $P_3(a)$,
$P_6(a)$, $\cdots$. As described above, our protocol is also secure
for noisy quantum channel.

So far we have proposed a QSDC protocol with pure entangled states
and analyzed the security for the present protocol. To check
eavesdropping in the transmission line, Alice inserts some decoy
photons in the transmitting photon sequence. After ensuring the
security of the quantum channel, Alice encodes her secret message on
the pure entangled states by using CNOT operation. The communication
parties measures each of their photons in $Z$-basis. Alice publishes
her measurement result and Bob can then recover Alice's secret
message. The present protocol is efficient in that all pure
entangled states are used to transmit the secret message. We also
point out that our protocol is secure with noisy quantum channel. As
for the experimental feasibility, our protocol can be realized with
today's technologies.

%%%%%%%%%%%%%%%%%%%%%%%%%%%%%%%%%%%%%%%%%%%%%%%%%%%%%%%%%%%%%%%%%%%%%%

\begin{acknowledgments} This work is supported by the
National Natural Science Foundation of China under Grant No.
60472032.
\end{acknowledgments}

%%%%%%%%%%%%%%%%%%%%%%%%%%%%%%%%%%%%%%%%%%%%%%%%%%%%%%%%%%%%%%%%%%%%%
%
%
\end{document}